\documentclass[
superscriptaddress,
amsmath,amssymb,
aps,
prb,
twocolumn,
floatfix,
]{revtex4-2}

\UseRawInputEncoding
\usepackage{natbib}
\usepackage{graphicx}
\usepackage{dcolumn}
\usepackage{bm}
\usepackage{xcolor}

\begin{document}

\title{Temperature-dependent thermal transport of single molecular junctions from semi-classical Langevin molecular dynamics}

\author{Gen Li}
\affiliation{School of Physics and Wuhan National High Magnetic Field Center, Huazhong University of Science and Technology, Wuhan 430074, China}
\author{Bing-Zhong Hu}
\affiliation{School of Physics and Wuhan National High Magnetic Field Center, Huazhong University of Science and Technology, Wuhan 430074, China}
\author{Nuo Yang}
\email{nuo@hust.edu.cn}
\affiliation{State Key Laboratory of Cool Combustion, and School of Energy and Power Engineering, Huazhong University of Science and Technology, Wuhan 430074, China}
\author{Jing-Tao L\"u}
\email{jtlu@hust.edu.cn}
\affiliation{School of Physics and Wuhan National High Magnetic Field Center, Huazhong University of Science and Technology, Wuhan 430074, China}

\date{\today}

\begin{abstract}
Thermal conductance of single molecular junctions at room temperature has been measured recently using picowatt-resolution scanning probes. However, fully understanding 
thermal transport in a much wider temperature range is needed for the exploration of energy transfer at single-molecular limit and the development of single molecular devices. Here, employing a semi-classical Langevin molecular dynamics method, a comparative study is performed on the thermal transport of alkane chain between Au and graphene electrodes, respectively. We illustrate the different roles of quantum statistics and anharmonic interaction in the two types of junctions. For graphene junction, quantum statistics is essential at room temperature, while the anharmonic interaction is negligible. For Au junction, it is the other way. Our study paves the way for theoretically understanding thermal transport of realistic single molecular junctions in the full temperature range by including both quantum statistics and anharmonic interaction within one theoretical framework.
\end{abstract}


\maketitle

\section{INTRODUCTION}
Although tremendous progress has been made in the measurement and understanding of electric, thermoelectric and optoelectric transport properties of single molecular junctions (SMJs)\cite{Evers2019,xiang_molecular-scale_2016,xin_concepts_2019,aradhya_single-molecule_2013,galperin_molecular_2012,Dubi2011}, the measurement of heat transport at the single molecular level is much more challenging and has become possible only very recently\cite{Cui2019,Cui2017a,Cui2017,Mosso2017,Mosso2019,Mosso2019a,Segal2016}. 
The dominant heat carriers are electrons in a metallic atomic wire, and the thermal conductance quantum has been observed\cite{Cui2017a,Mosso2019a}. However, in a single molecular junction, the dominant heat carriers become phonons due to the reduced electrical conductance\cite{Cui2019,Mosso2017,Mosso2019}. The length dependence of single molecular thermal conductance $\kappa$ has been studied, and the experimental results\cite{Cui2019} confirmed early theoretical prediction\cite{Segal2003,Li2015,klockner_length_2016}. 

So far, the experimental study has focused on $\kappa$ near room temperature (RT) $T=300$ K. To fully understand thermal transport properties of SMJs, the temperature dependence of $\kappa$ has to be studied, where theoretical methods applicable to the full temperature range is required. Commonly available methods have their limitations in this respect. Classical molecular dynamics (MD) can not be applied to low temperature regime, especially in the regime well below Debye temperature, where quantum statistics has to be taken into account. While the nonequilibrium Green's function (NEGF) method is fully quantum mechanical, it is not easy to deal with anharmonic interaction in the high temperature regime\cite{Wang2007}. Consequently, it is commonly used within the harmonic approximation\cite{lu2008,Li2015,Klockner2019,martinez2019}. A general theoretical method applicable in the full temperature range is a prerequisite to understand temperature dependent thermal transport in SMJs.

In this work, we employ a semi-classical Langevin molecular dynamics (SCLMD) method\cite{Wang2008,Dammak2009,Ceriotti2009,Barrat2011,Lu2012,Stella2014,Ness2014,Ness2016,Shen2016a,Kantorovich2016,Carpio-Martinez2021,Kantorovich2008,Kantorovich2008a,Lu2019,MoghaddasiFereidani2019,Lu2020} to study the temperature dependent $\kappa$ of single molecular junctions. The statistical properties of the Langevin thermal baths are treated quantum-mechanically, and the deterministic Hamiltonian dynamics of the system is treated classically. It has been shown that this semi-classical approach is asymptotically correct at both low and high temperature limits\cite{Wang2007,Dammak2009,Ceriotti2009}. 
We consider the prototypical alkane chain between two types of electrodes, Au and graphene, respectively. These two electrode materials have very different phonon bandwidth and Debye temperature (165 K for Au\cite{Neighbours1958,Balerna1986} and 2100 K for graphene\cite{Tohei2006,Tewary2009}). We show that at RT quantum statistics is essential for graphene junction, but not for Au junction. Meanwhile, anharmonic effect already shows up at RT in Au junctions and gives rise to $\sim 15\%$ reduction of $\kappa$ compared to the harmonic case. These results extend our understanding of heat transport in SMJs to a much wider temperature range and are helpful for the study of thermoelectricity, current-induced heating in SMJs\cite{Dubi2011,Cui2017,Lu2019}.

\begin{figure}[hbt!]
    \includegraphics[width=0.9\columnwidth]{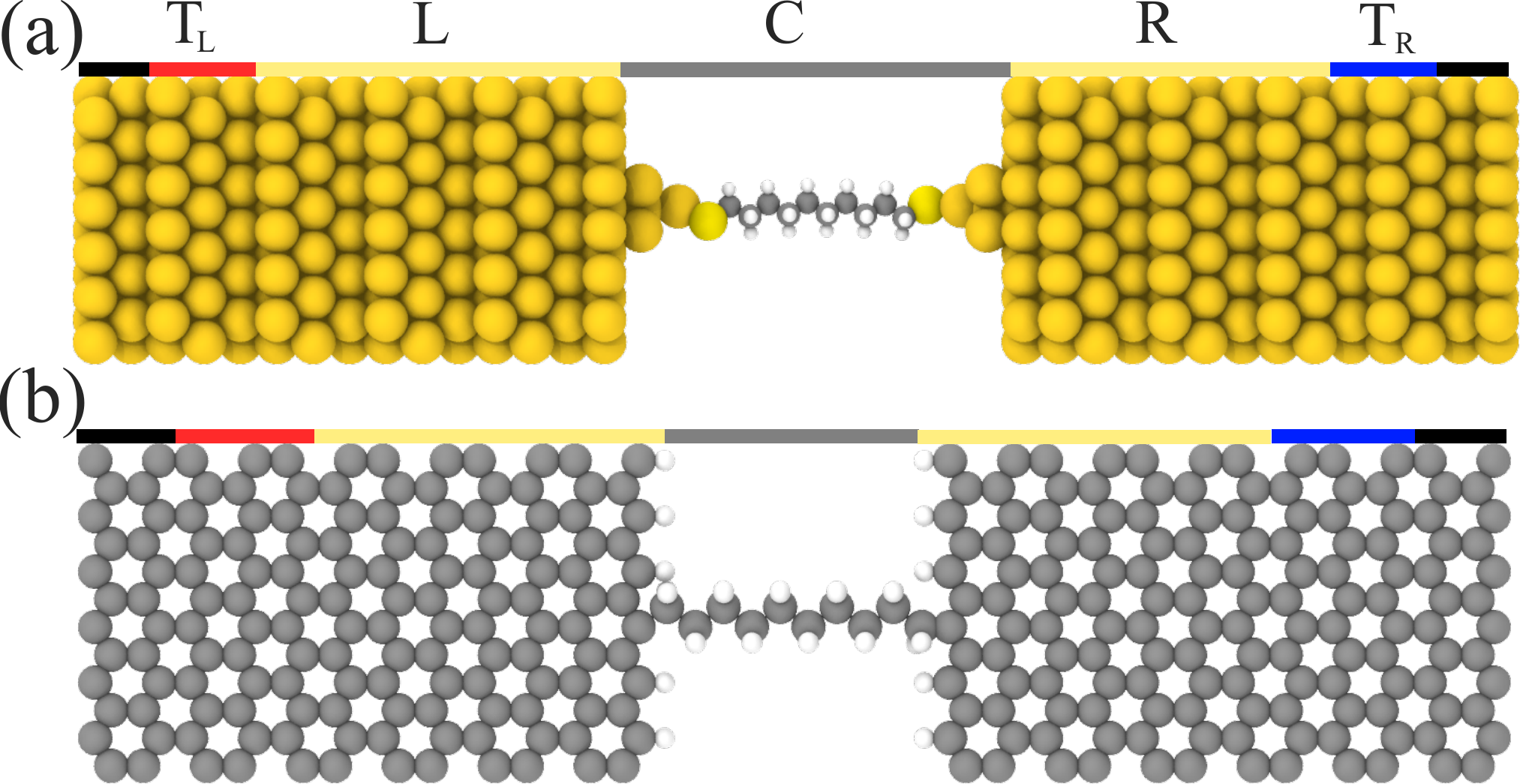}
    \caption{Alkane chain consisting of $N$ CH$_2$ units placed between Au (a) and graphene (b) electrodes. Atoms under the black lines are fixed, and the rest atoms are dynamical. Those under the red and blue lines are connected to quantum Langevin thermal baths at high and low temperatures, respectively. OVITO is used for visualization\cite{ovito}. }
    \label{fig:structure}
\end{figure}

\section{METHODS}

\subsection{Semi-classical Langevin molecular dynamics}
The semi-classical Langevin equation is written as\cite{Wang2008,Dammak2009,Ceriotti2009,Barrat2011,Lu2012,Stella2014,Ness2014,Ness2016,Shen2016a,Kantorovich2016,Carpio-Martinez2021}
\begin{equation}
\ddot{\mathbf{u}} = -\frac{\partial}{\partial \textbf{u}}V - \gamma \dot{\textbf{u}} + \textbf{f}.
\end{equation}
Here, $\mathbf{u}$ is a vector that contains the mass-normalized displacement from the corresponding equilibrium position of each atomic degree of freedom (DOF), i.e., $u_i = \sqrt{m_i}(R_i-R_i^{(0)})$, where $m_i$, $R_i$ and $R_i^{(0)}$ are the mass, time-dependent and equilibrium position of the $i$-th DOF, respectively. The first term on the right hand side represents the potential force, the second term is the friction, and the last term is the fluctuating force. The last two terms are from the quantum Langevin baths and apply only to the DOF directly connected to the two thermal baths (atoms under red and blue lines in Fig.~\ref{fig:structure}). In general, the friction term depends on the history of velocity, leading to a time convolution with a memory kernel. Here, we use the time local version. The memory effect can be taken into account by including extra atoms in the electrodes explicitly in the MD simulation\cite{Stella2014,Ness2014}. Moreover, we take the same friction coefficient $\gamma$ for all the DOF. The quantum effect is introduced by the fluctuating force with a colored noise spectrum
\begin{equation}
 S_{ij}(\omega) = 2\hbar\omega\gamma \delta_{i,j} \left[n_B(\omega,T)+\frac{1}{2}\right],
 \label{eq:sijq}
\end{equation}
with 
\begin{equation}
n_B(\omega,T)=\frac{1}{{\rm exp}(\hbar\omega/k_BT)-1}
\end{equation}
the Bose-Einstein distribution function, and $k_B$ the Boltzmann constant, $T$ the absolute temperature. With this quantum statistics, it has been shown that quantum ballistic thermal transport can be reproduced by the semi-classical equation\cite{Wang2007}. 
At high temperature limit, Eq.~(\ref{eq:sijq}) reduces to
\begin{equation}
    S_{ij}(\omega) = 2 k_BT\gamma \delta_{i,j},
\end{equation}
and we recover the classical Langevin equation.

The potential force is obtained by linking the homemade MD program with LAMMPS\cite{Plimpton1995}. For the junction with graphene electrodes, the second generation reactive empirical bond order (REBO) potential\cite{Brenner2002} is used, while for the junction with Au electrodes, a reactive force field (ReaxFF)\cite{Senftle2016,Jarvi2011,Aktulga2012,Bae2013} optimized for Au-S-C-H systems is used. Both potentials have been used to calculate $\kappa$ of similar junctions\cite{Noshin2017,DiPierro2018,Klockner2019,Diao2020}.

\begin{figure}
    \includegraphics[width=0.9\columnwidth]{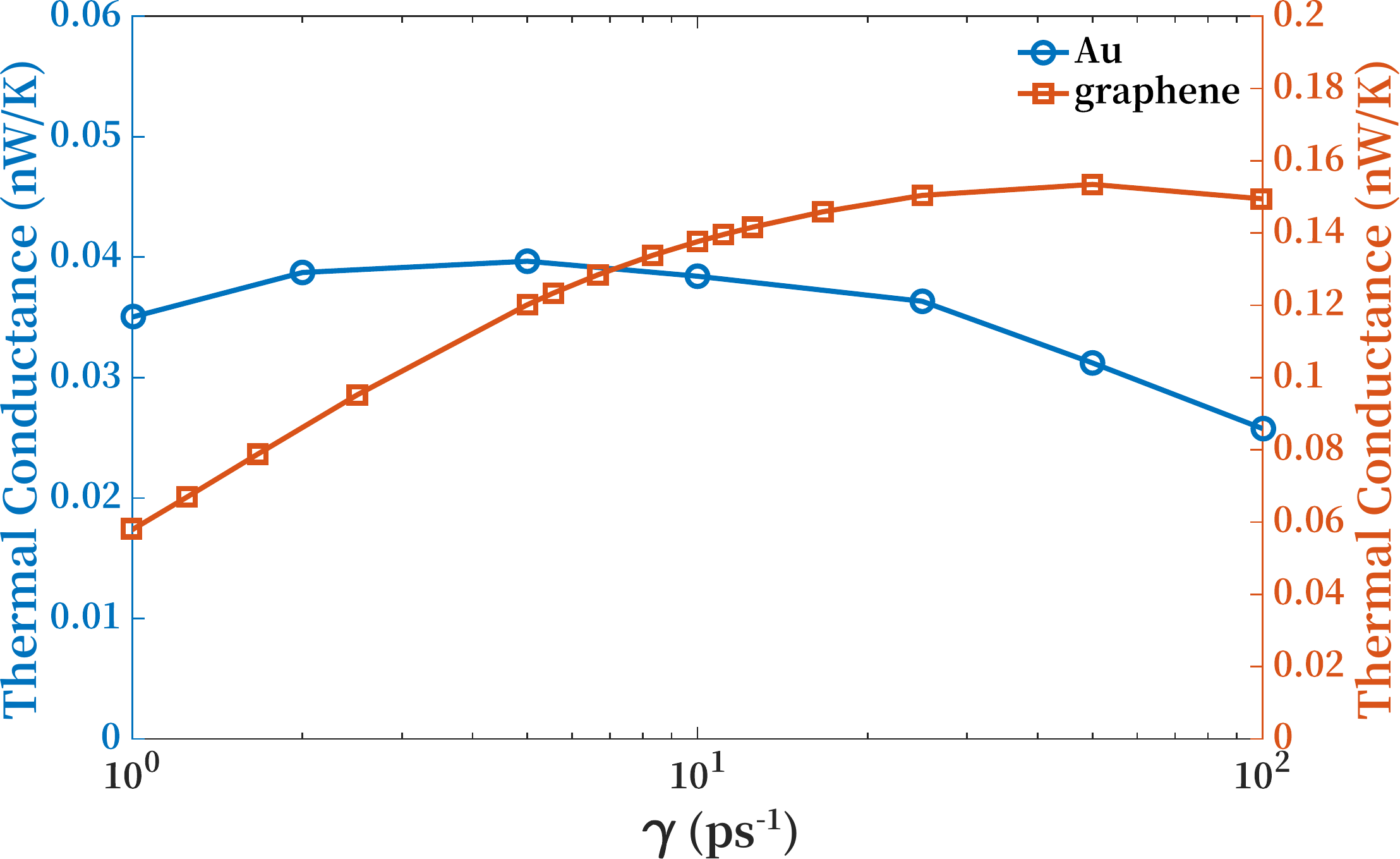}
    \label{fig-s-damp-au}
    \caption{
    Dependence of thermal conductance at room temperature on the friction coefficient $\gamma$. The results are obtained from NEGF using harmonic force constants. The molecule has 10 CH$_2$ units. According to these results, We have chosen $\gamma =10$ ps$^{-1}$ for all the following results reported in this work.  }
    \label{fig:damp}
\end{figure}


In the MD simulation, the whole system is divided into several parts (Fig.~\ref{fig:structure}): the central junction, the thermal bath region and the fixed region. The friction and the fluctuating force act on atoms in the thermal bath region only (atoms under blue and red lines in Fig.~\ref{fig:structure}). 
Periodic boundary condition is used in the direction perpendicular to the transport direction. A symmetric temperature difference of $\delta = 0.1 T$ is applied between the two thermal baths, with $T$ the average temperature.
Time steps $dt = 1$ fs and $dt = 0.5$ fs are respectively used for the alkane junction with Au and graphene electrodes in the MD simulation. The final results are obtained by averaging three independent MD runs, where each run lasts for $4\times 10^6$ steps\footnote{We have shared the structure of the molecular junctions and one typical MD trajectory at: https://github.com/sclmd/structures-trajectories . Other data related to this work is available upon request.}. We have used $\gamma = 10$ ps$^{-1}$ throughout (see Fig.~\ref{fig:damp} for the effect of $\gamma$ on $\kappa$), and
have further ignored the zero point energy in the power spectrum for better convergence. Comparison of results with and without zero point energy can be found in Fig.~\ref{fig-q-tb-all} (b). 
The kinetic energy distribution of the system can be studied using the power spectrum, which is defined by
\begin{equation}
    C_{vv}(\omega) = \sum_i C_{v_i v_i}(\omega) = \sum_i \int dt C_{v_i v_i}(t) e^{i\omega t},
\end{equation}
where
\begin{equation}
    C_{v_i v_i}(t) = \langle v_i(t) v_i(0) \rangle = \frac{1}{t_0}\int dt' v_i(t'+t)v_i(t')
\end{equation}
is the velocity correlation function at steady state. The time integration is over the full simulation time $t_0$. 

\subsection{Harmonic calculation using NEGF}
To compare with the MD results, we also calculate $\kappa$ within the harmonic approximation using the NEGF method. The harmonic force constants are obtained from the same potential as MD. The Landauer formula is used to calculate $\kappa$\cite{Wang2008,lu2008}
\begin{equation}
    \kappa = \int_0^{+\infty}\frac{d\omega}{2\pi} \mathcal{T}(\omega) \frac{\partial n_B}{\partial T},
    \label{eq:negfk}
\end{equation}
with the transmission coefficient given by the Caroli formula
\begin{equation}
    \mathcal{T}(\omega) = {\rm Tr}[D^r(\omega) \Gamma_L(\omega)D^a(\omega)\Gamma_R(\omega)],
\end{equation}
where $D^r(\omega)$ and $D^a(\omega)$ with $D^r(\omega)=(D^a)^\dagger(\omega)$ are the phonon retarded and advanced Green's functions, $\Gamma_\alpha = i[\Pi_\alpha(\omega)-\Pi_\alpha^\dagger(\omega)]$ the coupling function to the bath $\alpha$, and $\Pi^{r/a}_\alpha(\omega)$ is the phonon retarded/advanced self-energy due to coupling to bath $\alpha$. 

\section{RESULTS AND DISCUSSIONS}

Previously, both the NEGF and classical MD (CMD) have been used to study thermal transport in SMJs\cite{lu2008,Li2015,Li2017,klockner_length_2016,Klockner2017,Sevincli2019}. However, both methods have their shortcomings. In practice, it is not easy to include the anharmonic phonon interaction in the NEGF calculation. Consequently, NEGF method is often used under harmonic approximation\cite{lu2008,Li2015,Klockner2019}. On the other hand, CMD largely overestimates the thermal conductance at low temperatures due to the classical Maxwell-Boltzmann statistics. Thus, neither method can be used to study the full temperature dependence of $\kappa$ without $a$ $prior$ analysis of the system. 

The advantage of the SCLMD is that it is asymptotically correct in both high and low temperature limits. In the high temperature limit, quantum effect is not important, and the anharmonic effect can be taken into account classically. In the low temperature limit, the anharmonic effect is not strong, and quantum statistics becomes crucial.  

\begin{figure}
	\includegraphics[width=0.9\columnwidth]{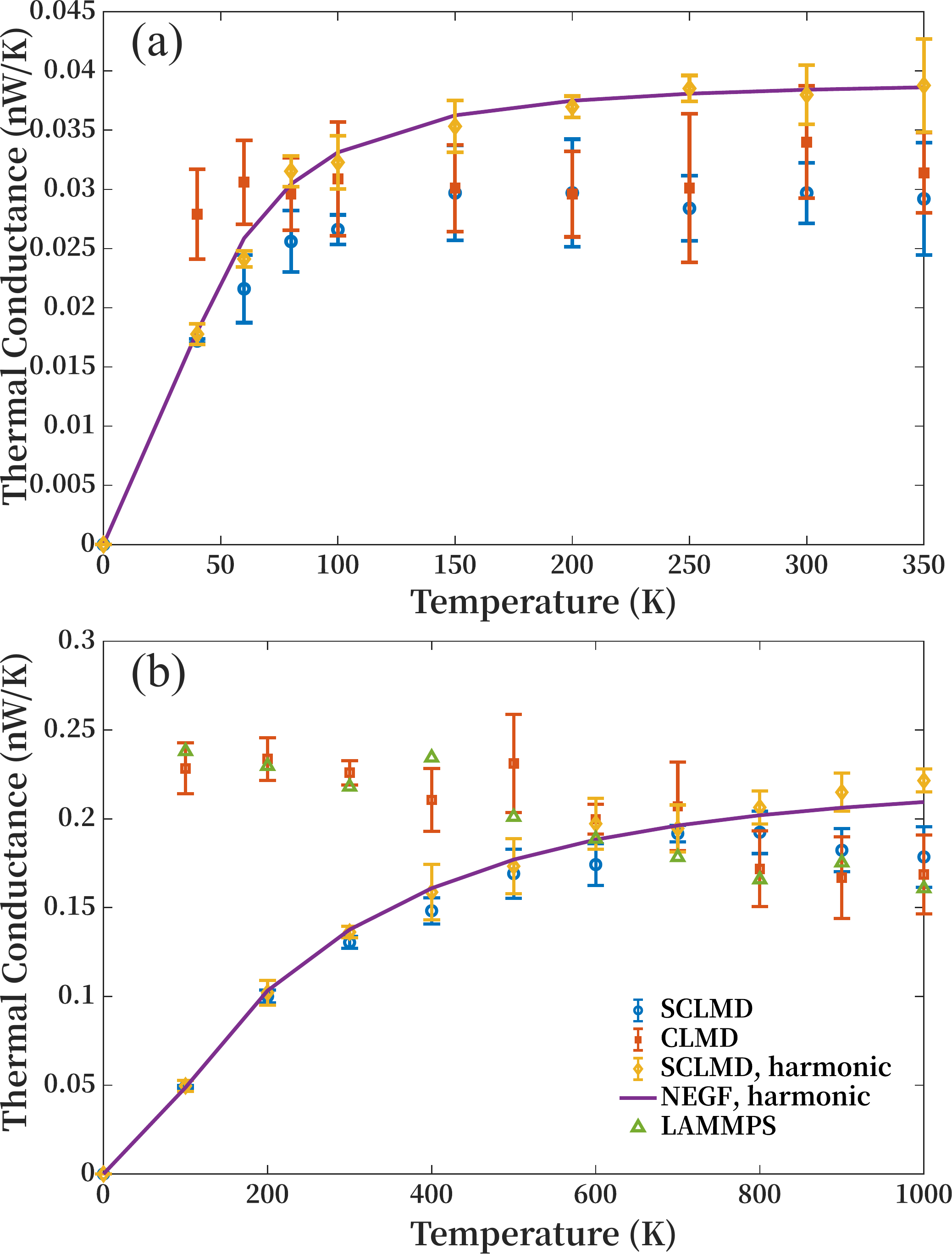}
	\caption{Temperature dependence of thermal conductance for molecular junction with 10 CH$_2$ units obtained from different approaches. (a) and (b) are for Au and graphene electrodes, respectively.}
	\label{fig:kt}
\end{figure}

We have shown that our homemade MD script correctly reproduces results in these two limits in Fig.~\ref{fig:kt}. Firstly,  we performed NEGF calculation using the harmonic force constants obtained from LAMMPS. It agrees well with our SCLMD results using the same set of harmonic forces. However, due to the limitation of classical statistics, the classical Langevin MD (CLMD) is not applicable in the low temperature regime, where the Bose-Einstein statistics can not be approximated by the classical Boltzmann statistics. This invalidates the use of CLMD at low temperature. Secondly, we compare results obtained from CLMD using LAMMPS and our homemade code. They agree with each other within the statistical error. Quantitatively, the average deviation between them, defined as $\kappa_r = (\kappa_{\rm LAMMPS}-\kappa_{\rm CLMD})/\kappa_{\rm CLMD}\times 100\%$ for each temperature,  is $\sim 3\%$.

We now discuss our main results shown in Fig.~\ref{fig:kt}. The temperature dependence of $\kappa$ obtained from SCLMD for the alkane junction with Au and graphene electrodes is shown in Fig.~\ref{fig:kt} (a) and (b), respectively.  
For the Au junction, SCLMD and CLMD results agree with each other at $T > 100$ K. The average relative difference for $T>100$ K, calculated as $\kappa_r = (\kappa_{\rm SCLMD}-\kappa_{\rm CLMD})/\kappa_{\rm CLMD}\times 100\%$ for each temperature, is $\sim 6\%$. We find that in this regime anharmonic effect reduces $\kappa$ by $\sim 15\%$ compared to the harmonic results. We attribute this reduction to the anharmonic interaction of vibrational modes in the molecule (see discussions below).  Quantum effect is important only for $T< 100$ K, where the SCLMD result deviates from that of CLMD. On the contrary, for graphene junction, quantum statistics is essential, leading to $\sim 20\%$ reduction of the thermal conductance at RT. The agreement between NEGF and SCLMD results indicates that the anharmonic effect does not influence the thermal conductance yet. 

\begin{figure}
	\includegraphics[width=0.9\columnwidth]{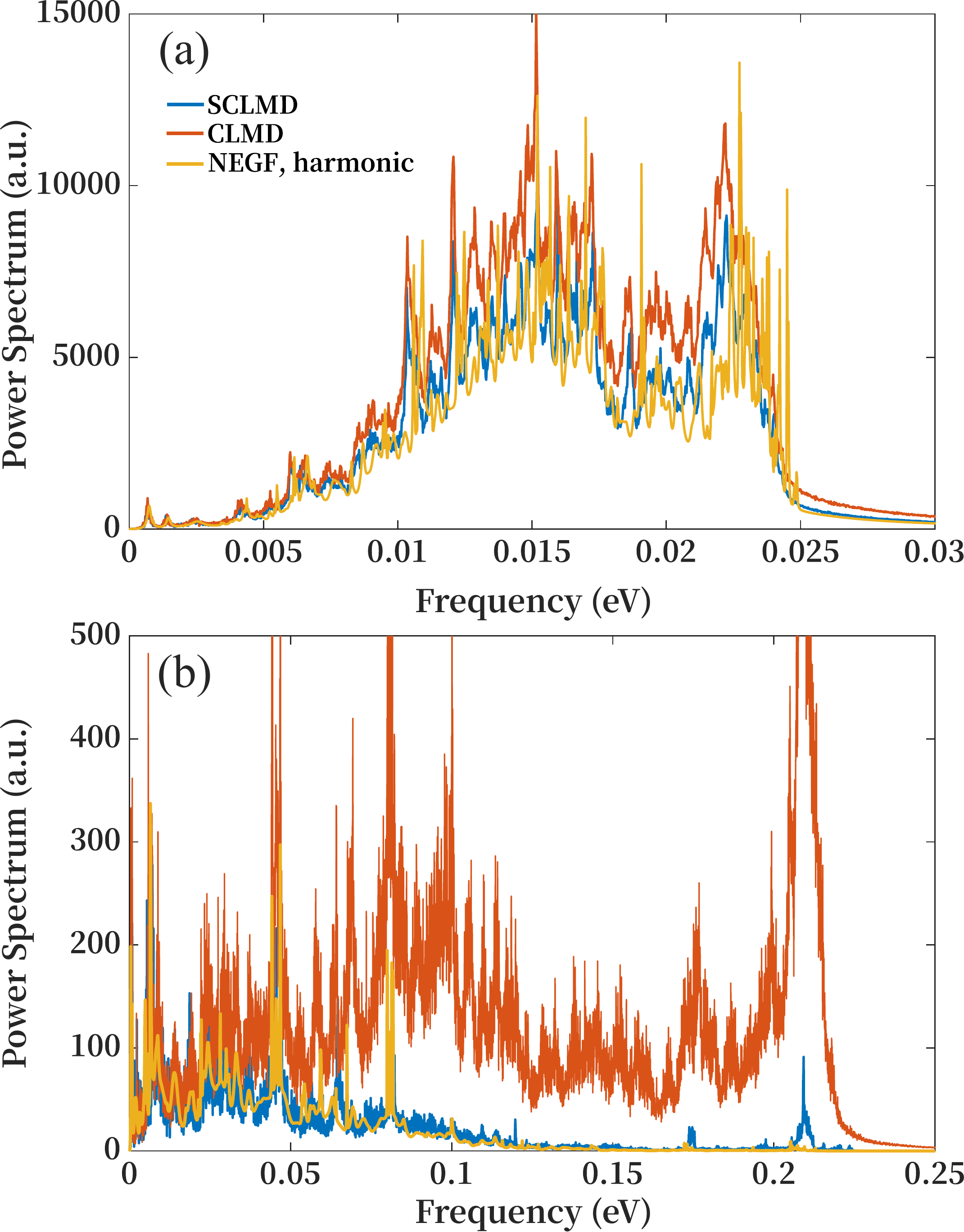}
	\caption{Power spectrum of the left electrode atoms obtained from different methods for Au (a) and graphene (b) electrodes. The average temperature is $T=300$ K. The power spectra of Au (a) obtained from the three methods are similar. However, for graphene junction (b), the CLMD, following Maxwell-Boltzmann statistics, gives much larger power spectrum at high frequency, leading to over-estimation of $\kappa$. The SCLMD overcomes this shortcoming by enforcing Bose-Einstein statistics in the Langevin baths. The results thus obtained are close to that of quantum-mechanical NEGF method. }
	\label{fig:power-leads}
\end{figure}

In Fig.~\ref{fig:power-leads} we plot the the power spectrum of electrode atoms at RT from SCLMD, CLMD and NEGF methods. For Au junction, their difference is small. However, for graphene junction, due to classical equipartition, the CLMD gives rise to much larger energy of the high frequency phonons. They furthermore contribute to thermal transport by coupling to the molecular vibrations, leading to overestimation of the thermal conductance. This unphysical effect is avoided in the SCLMD, where the phonons follow the quantum Bose-Einstein distribution. The power spectrum explains why CLMD can not be used to study RT $\kappa$ of graphene junction.

\begin{figure}[h!]
	\includegraphics[width=0.9\columnwidth]{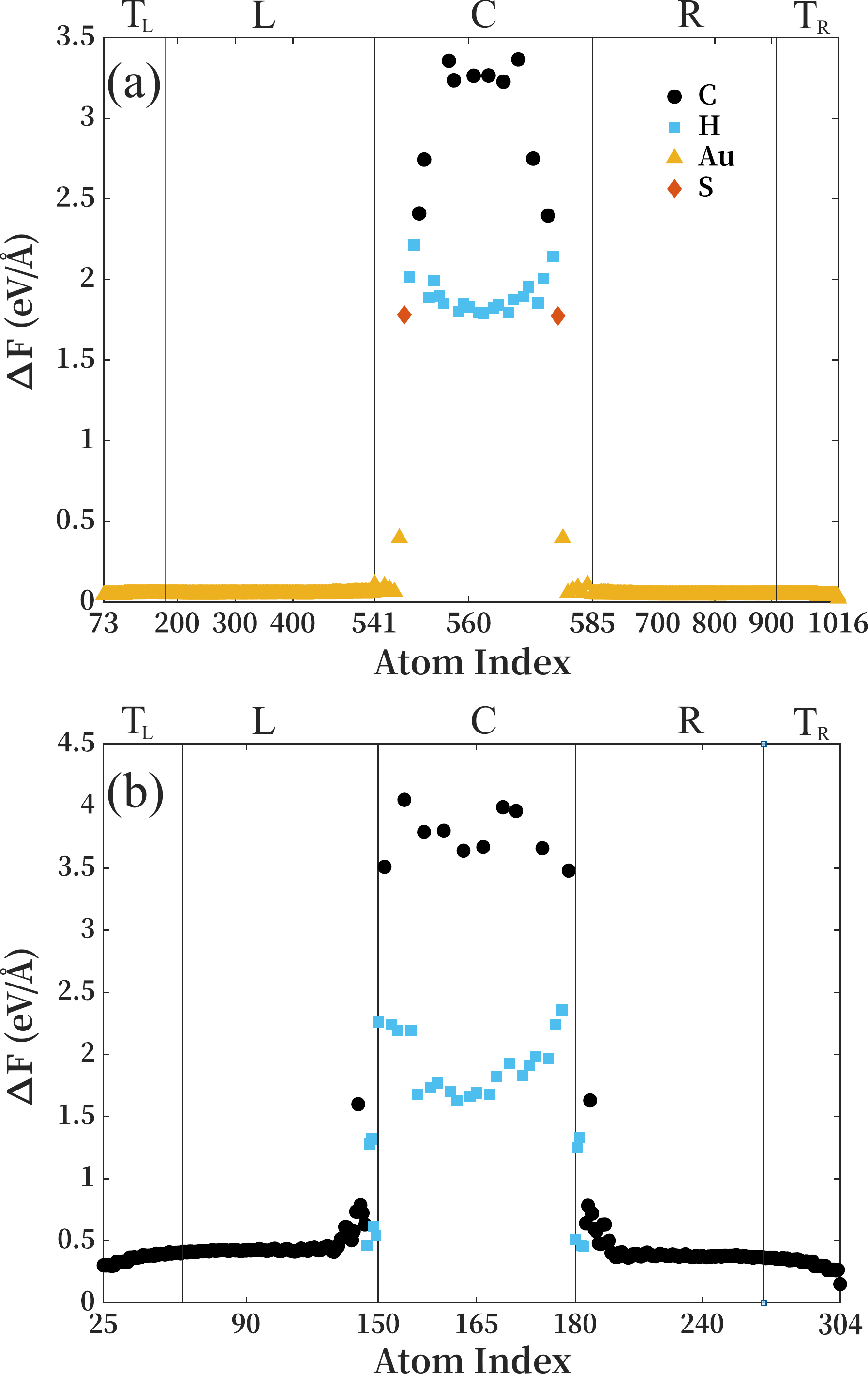}
	\caption{The average of absolute difference $\Delta F = |F_n-F_h|$ between the anharmonic force from empirical potential $F_n$ and harmonic force $F_h$ from the dynamic matrix for each atom along the harmonic trajectories. Average is done over  $2\times 10^5$ MD steps.}
	\label{fig-comp-hn}
\end{figure}

We notice the deviation between NEGF harmonic and the MD results in the high temperature regime ($T>100$ K for Au junction, $T>800 K$ for graphene junction). This is due to the anharmonic interaction between different phonon/vibrational modes.
In molecular junctions, the anharmonic interaction has two opposite effects on $\kappa$. Firstly, it may enhance thermal transport by redistributing heat between different modes, especially at the molecule-electrode interfaces. This opens heat transport channel of high frequency vibrations within the molecule that are out of the phonon band of Au electrodes. Secondly, it may cause scattering of the low frequency phonons that carry most of the heat. Comparison between the harmonic NEGF and SCLMD results suggests the latter is dominant here. This is also indicated from much larger deviation of the anharmonic forces from the harmonic ones in the molecule than in the electrodes (Fig.~\ref{fig-comp-hn}).

Comparing the two types of junctions, we find that $\kappa$ of the graphene junction is about one order of magnitude larger than that of the Au junction. This can be attributed to better phonon spectrum overlap between the molecule and the graphene electrodes. In biased SMJs, electron-vibration interaction results in Joule heating and current-induced forces in the molecule\cite{Dundas2009,Lu2010,Lu2015}. Larger $\kappa$ is favorable for efficient energy transport from the molecular vibrations to electrode phonons. Thus, single molecular junctions with graphene electrodes are promising candidates for constructing stable single molecular devices\cite{Jia2016}.

\begin{figure}[h!]
	\includegraphics[width=1.0\columnwidth]{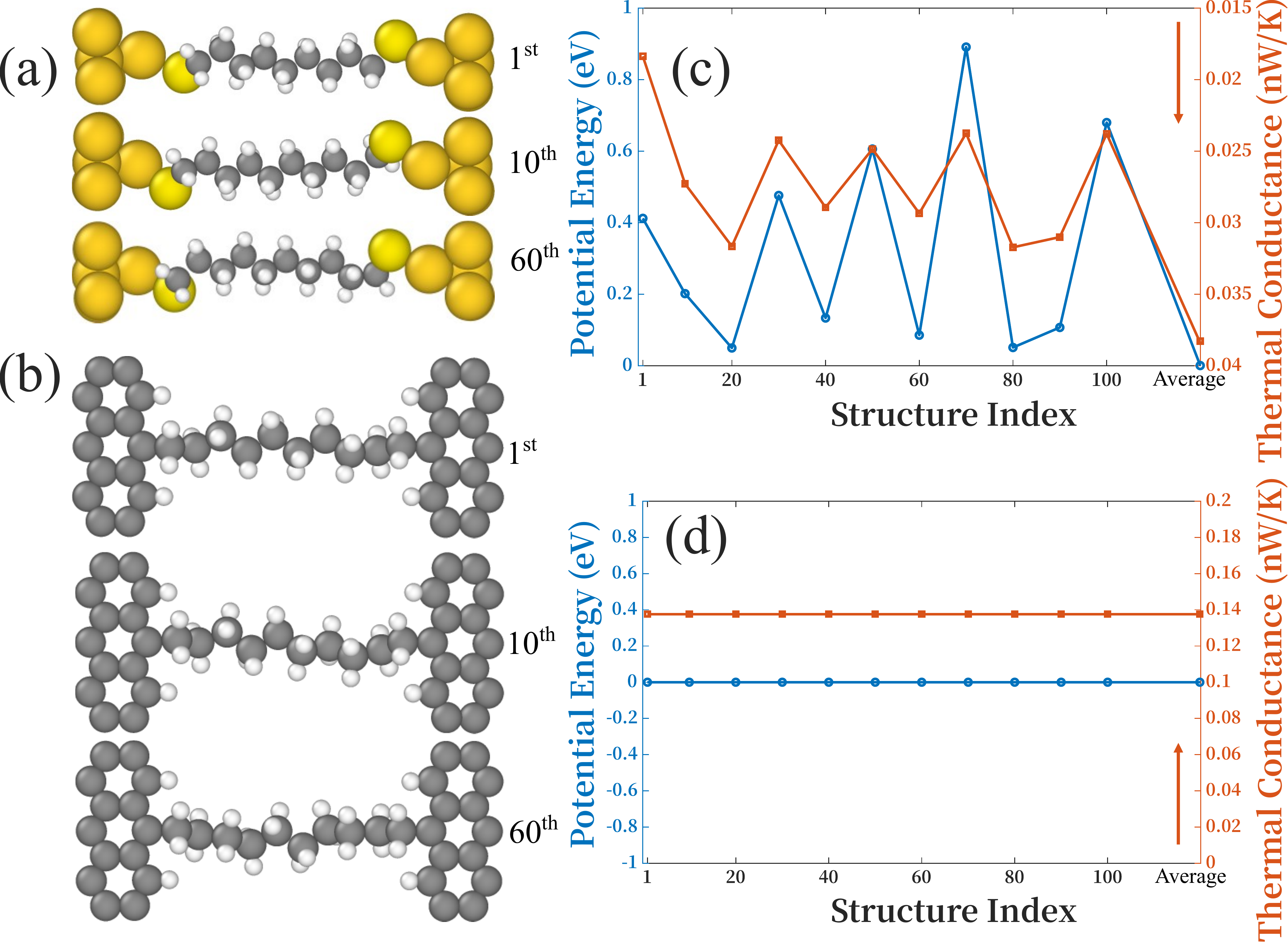}
	\caption{Typical structures extracted from the MD trajectory (a-b), their  thermal conductance (red)  at 300 K, and potential energy with respect to the lowest energy conformer used in Fig.~\ref{fig:kt} (blue). We have taken $\sim 100$ structures that deviate most from the lowest energy conformer. The deviation is defined as $d=\sum_i \Delta R_{i}^2$, where $\Delta R_{i}$ is the displacement of atom $i$ from the lowest energy conformer. The 1st structure has the largest deviation. To show the correlation between potential energy and thermal conductance, we have reversed the direction of thermal conductance in (c) as indicated by the arrow.}
	\label{fig-structures}
\end{figure}

So far, we have only considered one conformer with the lowest energy for the two types of junctions. During the MD run, we actually generate many different structures. To investigate how $\kappa$ depends on the molecular conformation, we have taken typical conformers from the first $\sim 100$ structures that deviate most from the lowest energy conformer within one MD run ($4\times 10^6$ steps), whose potential energy is set to zero. Typical structures, their potential energy, thermal conductance are shown in Fig.~\ref{fig-structures}. For Au junction, the thermal conductance spans within 0.02-0.04 nW/K. There is a close correlation between potential energy and thermal conductance. Conformers with larger potential energy have smaller thermal conductance. For graphene junction, their potential energy and thermal conductance are almost the same as the lowest energy conformer.

\begin{figure}[h!]
	\includegraphics[width=0.9\columnwidth]{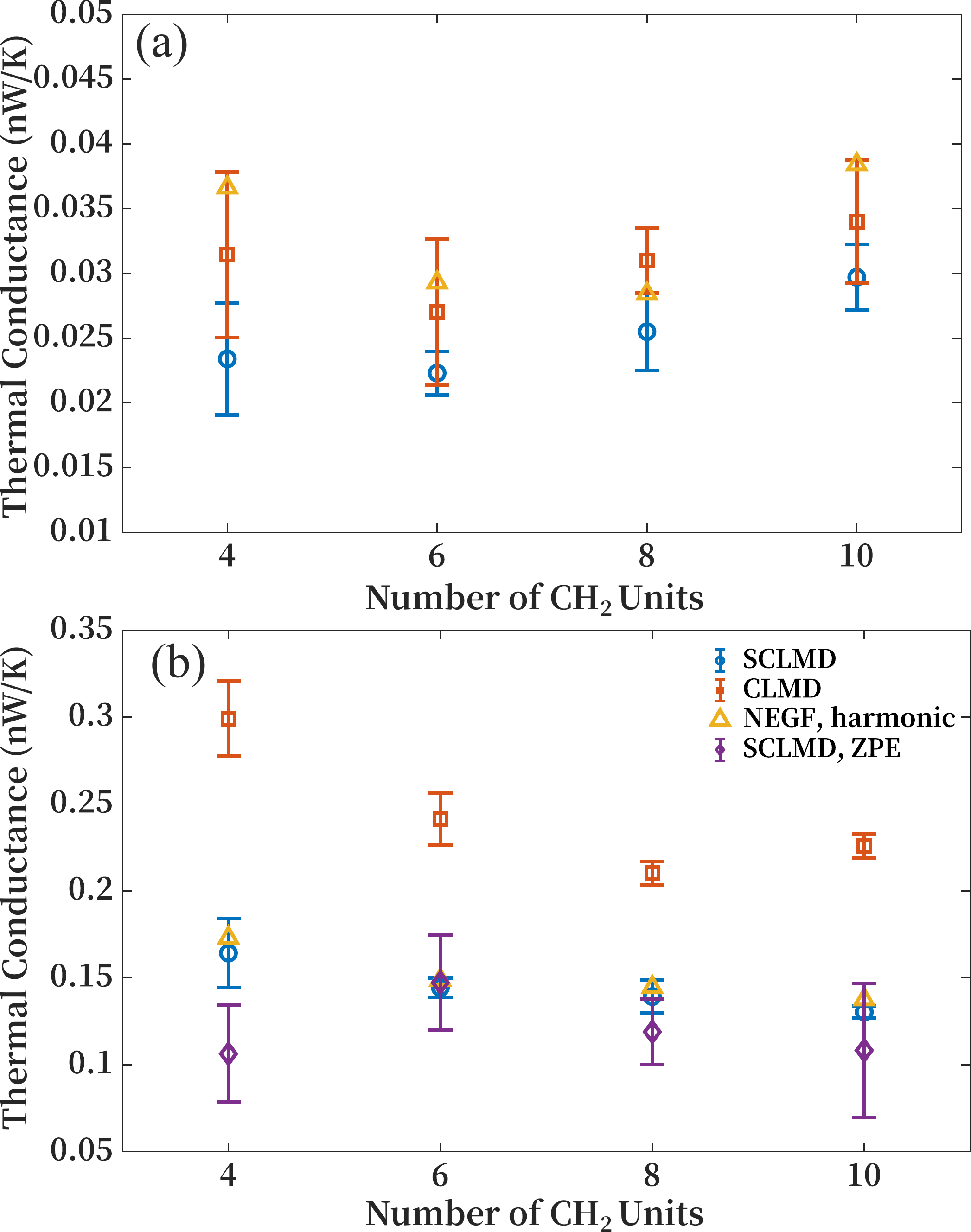}
	\caption{Variation of thermal conductance at 300 K with molecular length between Au (a) and graphene (b) electrodes. Different approaches give a similar trend in length dependence. But the CLMD overestimates the thermal conductance of graphene junction. Moreover, results with and without zero point energy (ZPE) are compared, showing that ZPE increases the uncertainty in the thermal conductance, but has small effect on the average.}
	\label{fig-q-tb-all}
\end{figure}

Length dependence of $\kappa$ for alkane chain has been studied intensively in the past years\cite{Segal2003,klockner_length_2016,Segal2016}. Theoretical study suggested a weak length dependence of $\kappa$, which was confirmed experimentally recently\cite{Cui2019}. In Fig.~\ref{fig-q-tb-all}, results obtained from the SCLMD and those from CLMD and NEGF approach for Au (a) and graphene (b) electrodes are compared, respectively. The weak length dependence is reproduced by all methods for both types of electrodes. For Au junction, the SCLMD results are consistently smaller than the NEGF results for all lengths. As we have discussed, we attribute this to the anharmonic vibrational interaction within the molecule.
For the graphene junction, CLMD overestimates the absolute value of $\kappa$ for all cases. This again shows the importance of quantum statistics for quantitative analysis of $\kappa$ in single molecular junction with graphene electrodes. Inclusion of zero point energy introduces larger uncertainty in the resulting thermal conductance. It will require longer MD simulation to reduce the uncertainty. But the average thermal conductance does not change much, although we do observe slight decrease compared to the result without zero point energy.

\section{CONCLUSION}
We have implemented a semi-classical Langevin molecular dynamics approach to study the thermal transport properties of single molecular junctions. We have performed a comparison of this method to commonly used classical molecular dynamics and the nonequilibrium Green's function method. The results show that our semi-classical Langevin approach can be used both at low temperature regime where quantum statistics is essential and at high temperature regime where anharmonic interaction plays a role. Applying this method to two types of molecular junctions, we find that thermal conductance of molecular junction with graphene electrodes is one order of magnitude larger than that with Au electrodes. The high thermal conductance of graphene junction is helpful for transport of excess heat from the central junction to the surrounding phonon environment in a working single-molecular optoelectronic device. For graphene electrodes, quantum statistics is important even at room temperature. The semi-classical Langevin approach provides an efficient and practical method for the theoretical study of single molecular thermal transport in the full temperature range.

\begin{acknowledgments}
This work was supported by the National Natural Science Foundation of China (Grant No. 21873033), the National Key Research and Development Program of China (Grants No. 2017YFA0403501 and No. 2018YFE0127800), Fundamental Research Funds for the Central Universities (Grant No. 2019kfyRCPY045), and the Program for HUST Academic Frontier Youth Team. We thank the National Supercomputing Center in Shanghai for providing computational resources.
\end{acknowledgments}

\bibliography{ref}


\end{document}